\documentstyle[epsfig]{aipproc}

\begin{document}

{\footnotesize \it \noindent 
   Invited talk, Proc.\ 5th Compton Symposium (Portsmouth NH, September 1999)}

\title{Diffuse Galactic Continuum \\ Gamma Rays}

\author{Andrew W. Strong$^*$, Igor V. Moskalenko$^{*\dagger\ddagger}$, 
   and Olaf Reimer$^*$}
\address{$^*$Max-Planck Institut f\"ur extraterrestrische Physik,
   85740 Garching, Germany\\
$^{\dagger}$Institute for Nuclear Physics, M.V. Lomonosov Moscow State 
   University,  Moscow, Russia\\ 
$^{\ddagger}$LHEA NASA/GSFC Code 660, Greenbelt, MD 20771, USA
}

\maketitle

\begin{abstract}
Galactic diffuse continuum $\gamma$-ray emission is intricately related to
cosmic-ray physics and radio astronomy.  We describe recent results from an
approach which endeavours to take advantage of this.  Information from
cosmic-ray composition constrains the propagation of cosmic rays; this in turn
can be used as input for $\gamma$-ray models.  The GeV $\gamma$-ray excess
cannot be explained as $\pi^o$-decay resulting from a hard nucleon spectrum
without violating antiproton and positron data; the best explanation at present
appears to be inverse-Compton emission from a hard interstellar electron
spectrum.  One consequence is an increased importance of Galactic inverse
Compton for estimates of the extragalactic background.  At low energies, an
additional point-source component of $\gamma$-rays seems to be necessary.

\end{abstract}

\section{Introduction}

This paper discusses recent studies of the diffuse continuum emission and their
connection with cosmic-ray physics.  The basic question concerns the origin of
the intense continuum emission along the Galactic plane observed by EGRET,
COMPTEL and OSSE.  The answer is surprisingly uncertain.  A comprehensive review
can be found in \cite{Hunter_review}.  The present work uses observational
results given in \cite{Strong98,StrongMattox96,Kinzer}; new imaging and spectral
results from COMPTEL are reported in \cite{Bloemen2000}.  Most of the analysis
reported here is based on the modelling approach described in \cite{sm98,smr98}.
First we present some results from cosmic-ray isotopic composition which bear
directly on the $\gamma$-ray models.  We then discuss the problems which arise
when trying to fit the $\gamma$-ray spectrum, and present possible solutions,
both at high and low energies.  The low energy (1--30 MeV) situation is
addressed in more detail in \cite{sm2000}, and additional references can be
found at \cite{website}.

Our basic approach is to construct a unified model which is as far as possible
realistic, using information on the gas and radiation fields in the Galaxy, and
current ideas on cosmic-ray propagation, including possible reacceleration; we
use these to predict many different types of observations:  direct measurements
in the heliosphere of cosmic ray nuclear isotopes, antiprotons, positrons,
electrons; and astronomical measurements of $\gamma$-rays and synchrotron
radiation.  Any given model has to be tested against all of these data and it is
a challenge to find even one which is consistent with all observations.  In fact
we will show that the full range of observations can only be accomodated by
additional components such as $\gamma$-ray point sources and also differences
between local direct measurements and large-scale Galactic properties of cosmic
rays.

\section{Cosmic ray nucleons}

\begin{figure}[h] 
   \centerline{
            \epsfig{file=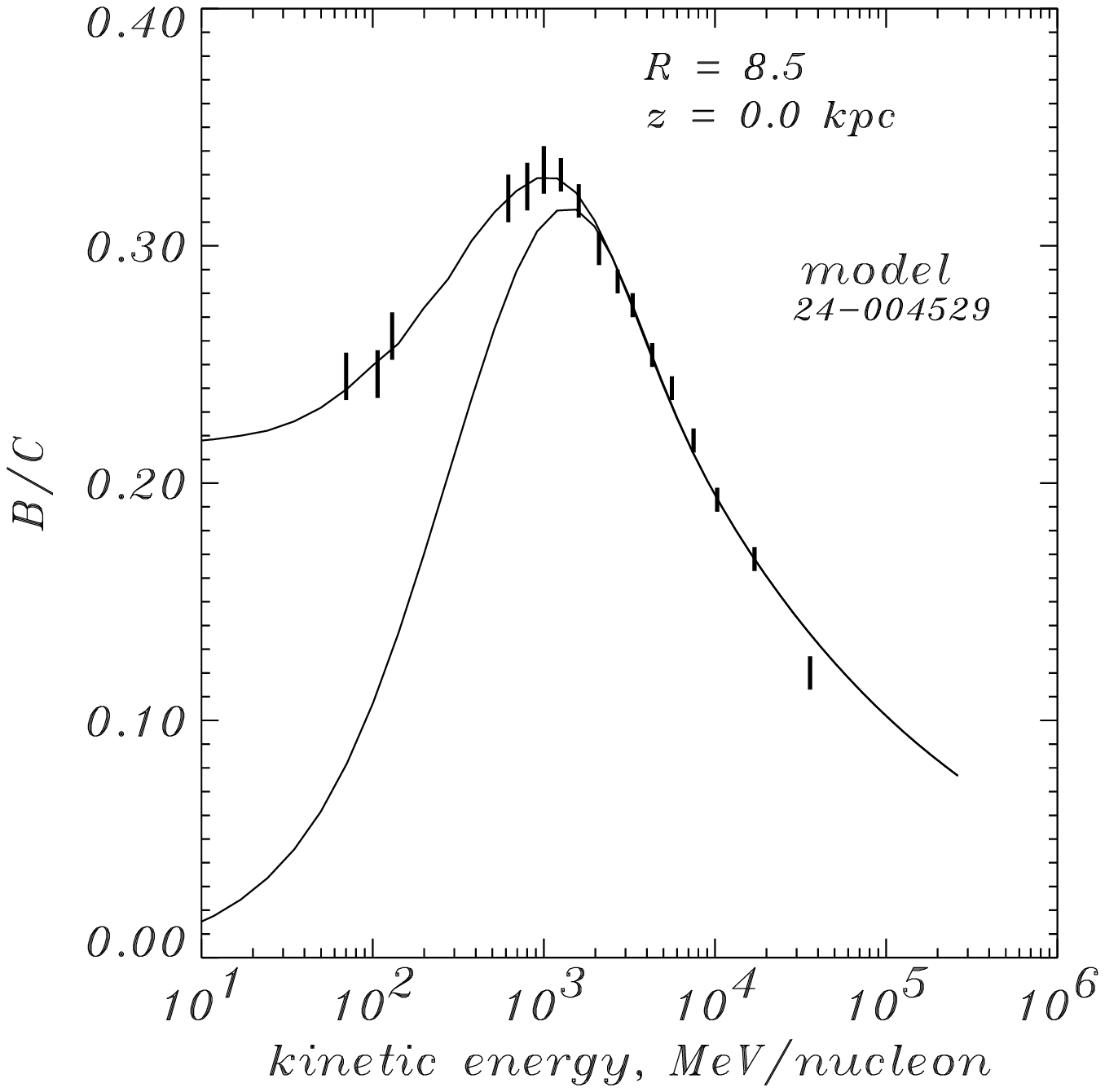,height=2.6in,width=2.8in}
            \epsfig{file=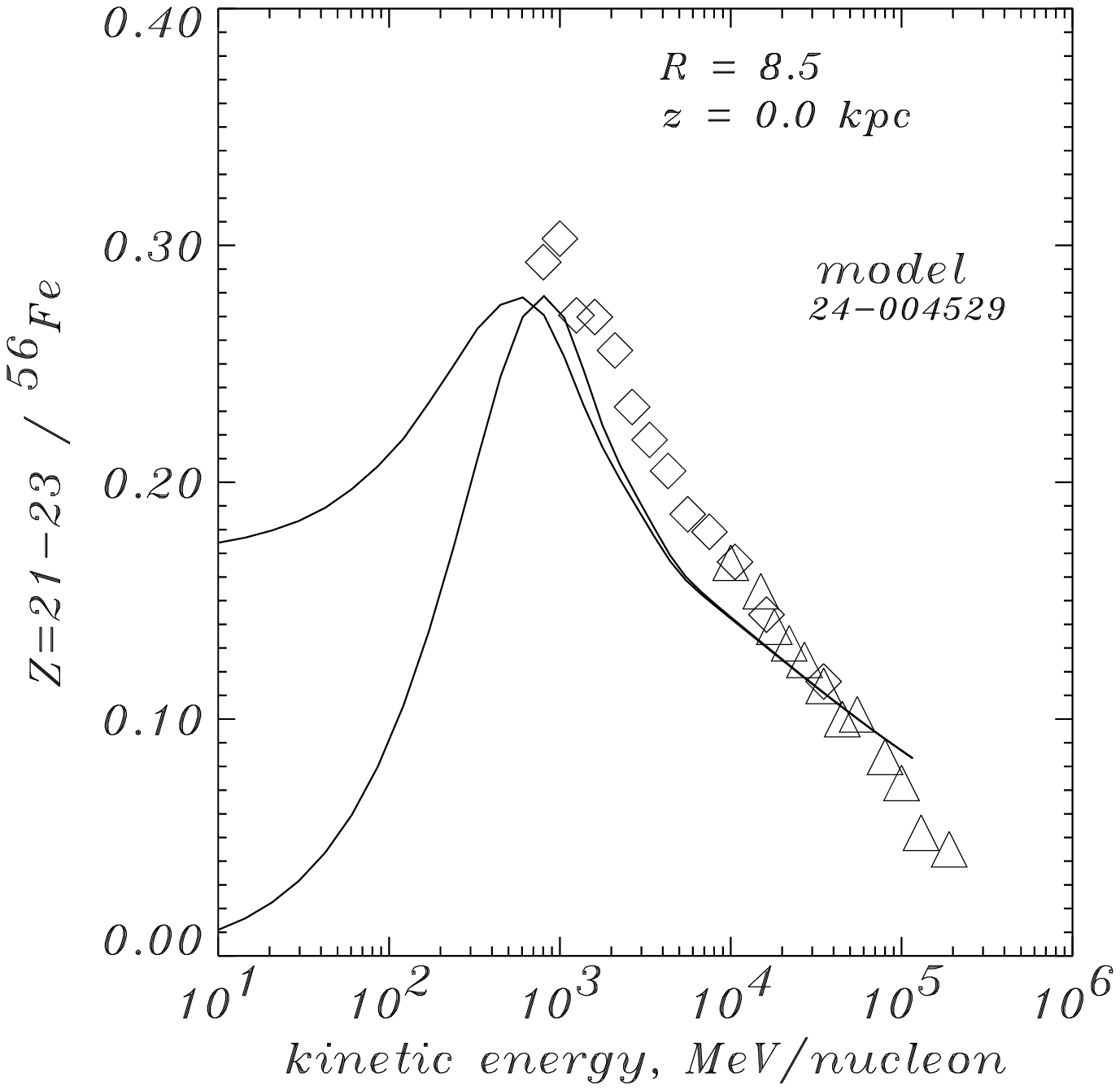,height=2.6in,width=2.8in}
   }
\vspace{10pt}
\caption{Cosmic-ray B/C and sub-Fe/Fe ratios for a diffusive halo model with 
reacceleration, halo height 4 kpc. For details of model and data see 
\protect\cite{sm99}. }
\label{BC}
\end{figure}

\begin{figure}[h] 
   \centerline{
            \epsfig{file=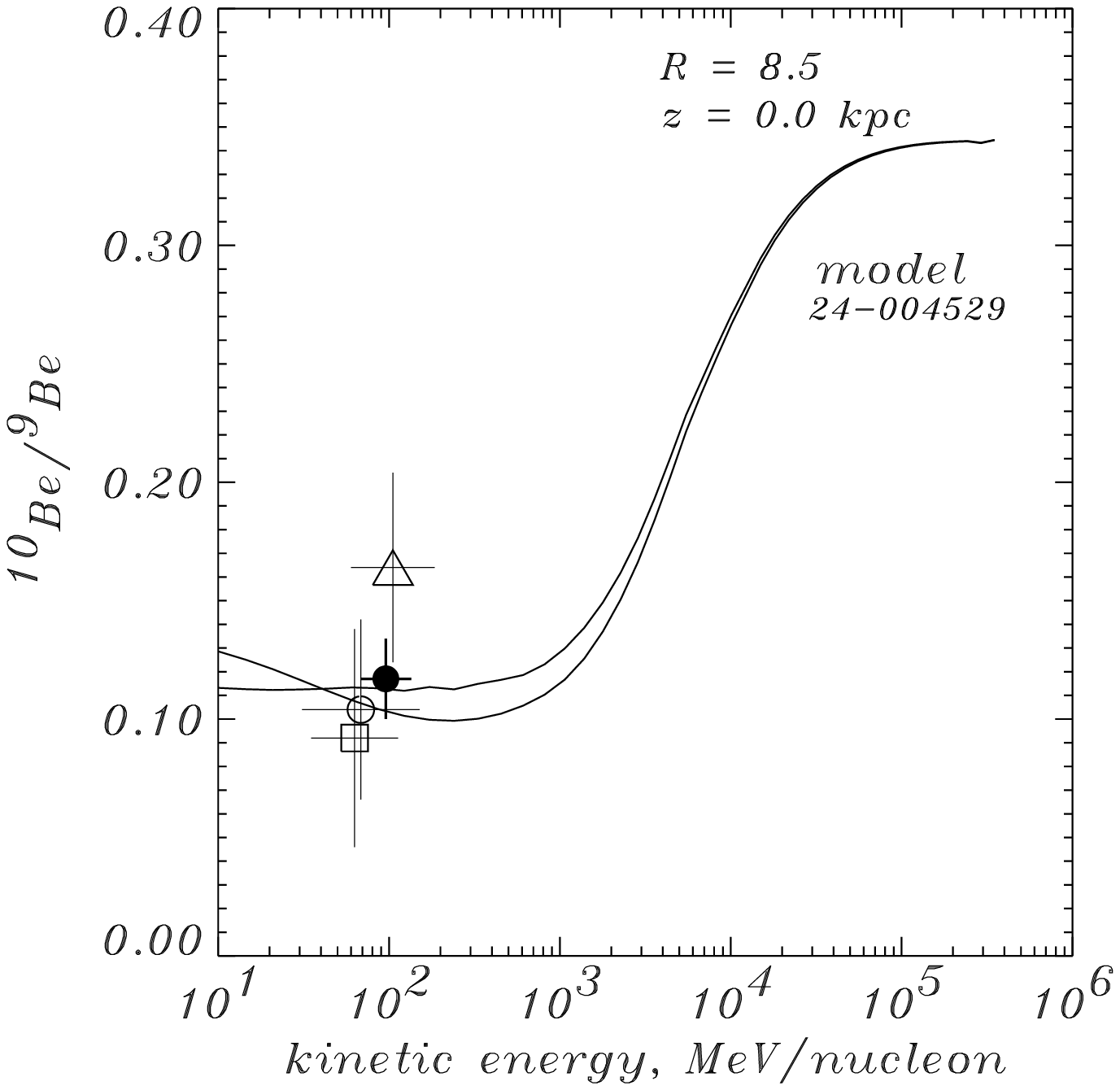,height=3.0in,width=2.5in}
            \epsfig{file=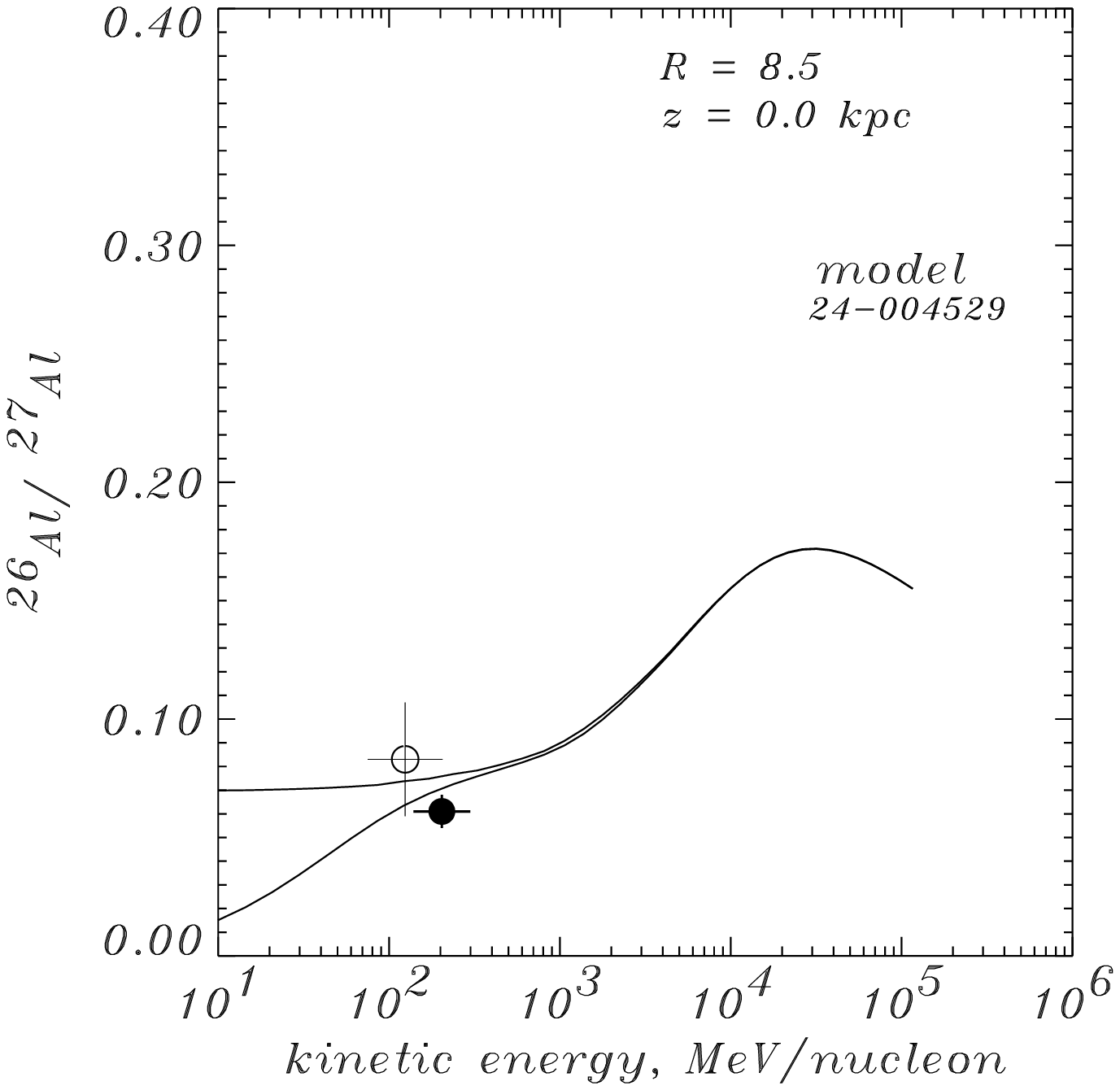,height=3.0in,width=2.5in}
   }
\vspace{10pt}
\caption{Cosmic-ray $^{10}$Be/$^{9}$Be  and $^{26}$Al/ $^{27}$Al ratio for the 
same model as used for Fig \protect\ref{BC}. For details of model and data see 
\protect\cite{sm99}.
}
\label{Be}
\end{figure}

First we show results from CR composition which are relevant to the propagation
of cosmic rays.  For a given halo size (defined here as the $z$ value at which
the cosmic-ray density goes essentially to zero) the parameters of the
diffusion/reacceleration model can be adjusted to fit the important
secondary/primary ratios, illustrated in Fig \ref{BC} for a halo size of 4 kpc.
In addition we can use the constraints on the halo size given by the radioactive
CR species $^{10}$Be and $^{26}$Al, Fig \ref{Be}.  For details of Ulysses
results on radioactive nuclei see
\cite{Connell2000,Connell98a,Connell98b,Simpson98}.  Based on Ulysses $^{10}$Be
data, a range for the halo height of 4--10 kpc was derived in \cite{sm98,sm99}.
This is consistent with other analyses \cite{Ptuskin98,Webber98}.  New results
from the Advanced Composition Explorer satellite (ACE) will constrain the halo
size better, but the above range is consistent with ACE results as presented in
\cite{ACE}.  Other radioactive nuclei ($^{36}$Cl and $^{54}$Mn) will provide
further independent information; at present one can only say that they are
consistent with the other nuclei.  Having obtained sets of propagation
parameters based on isotopic composition, we can proceed to use the model to
study diffuse $\gamma$-rays.

\section{Gamma rays}

\begin{figure}[b!] 
   \centerline{
      \epsfig{file=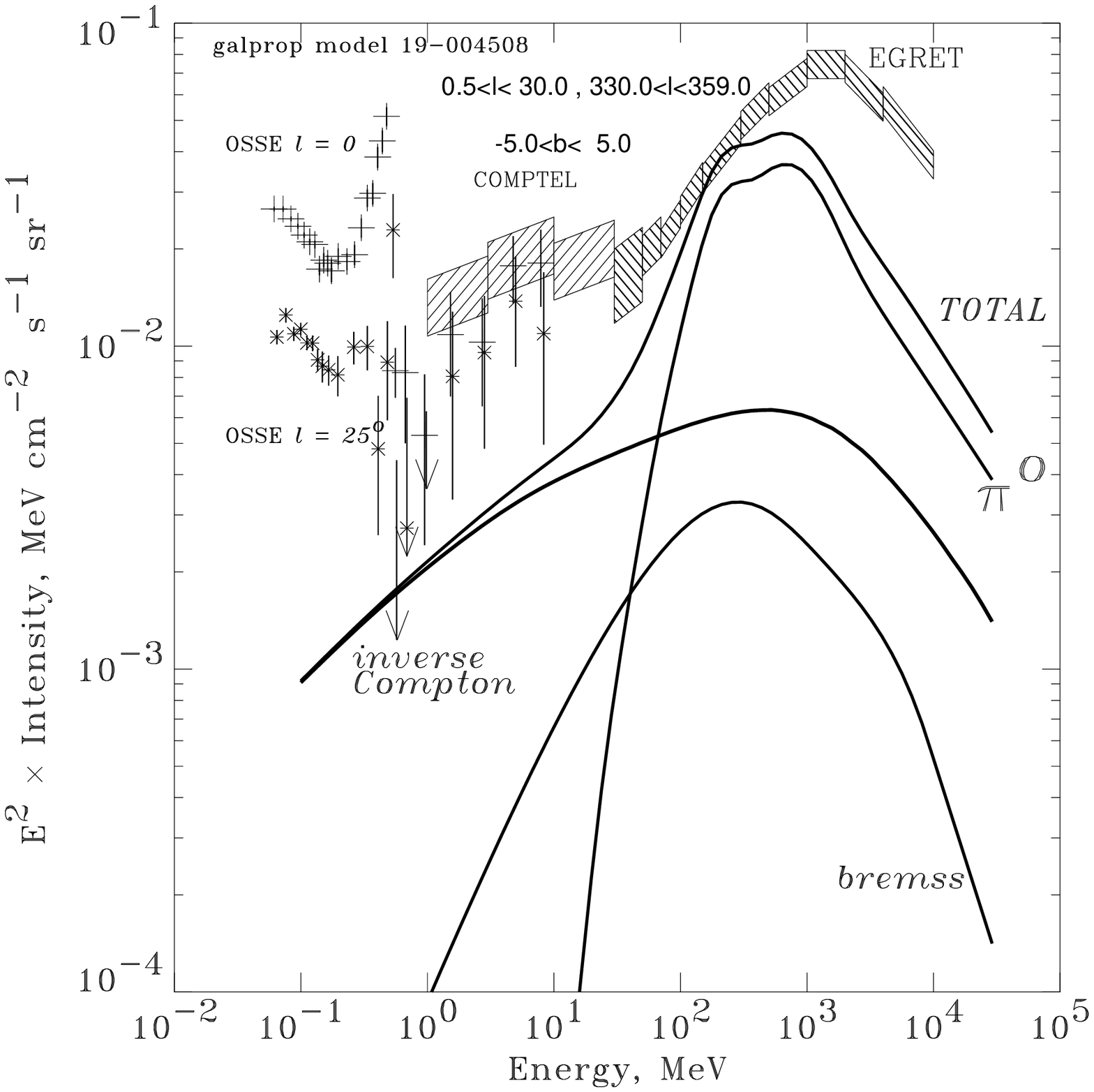,height=3.0in,width=2.8in}
      \epsfig{file=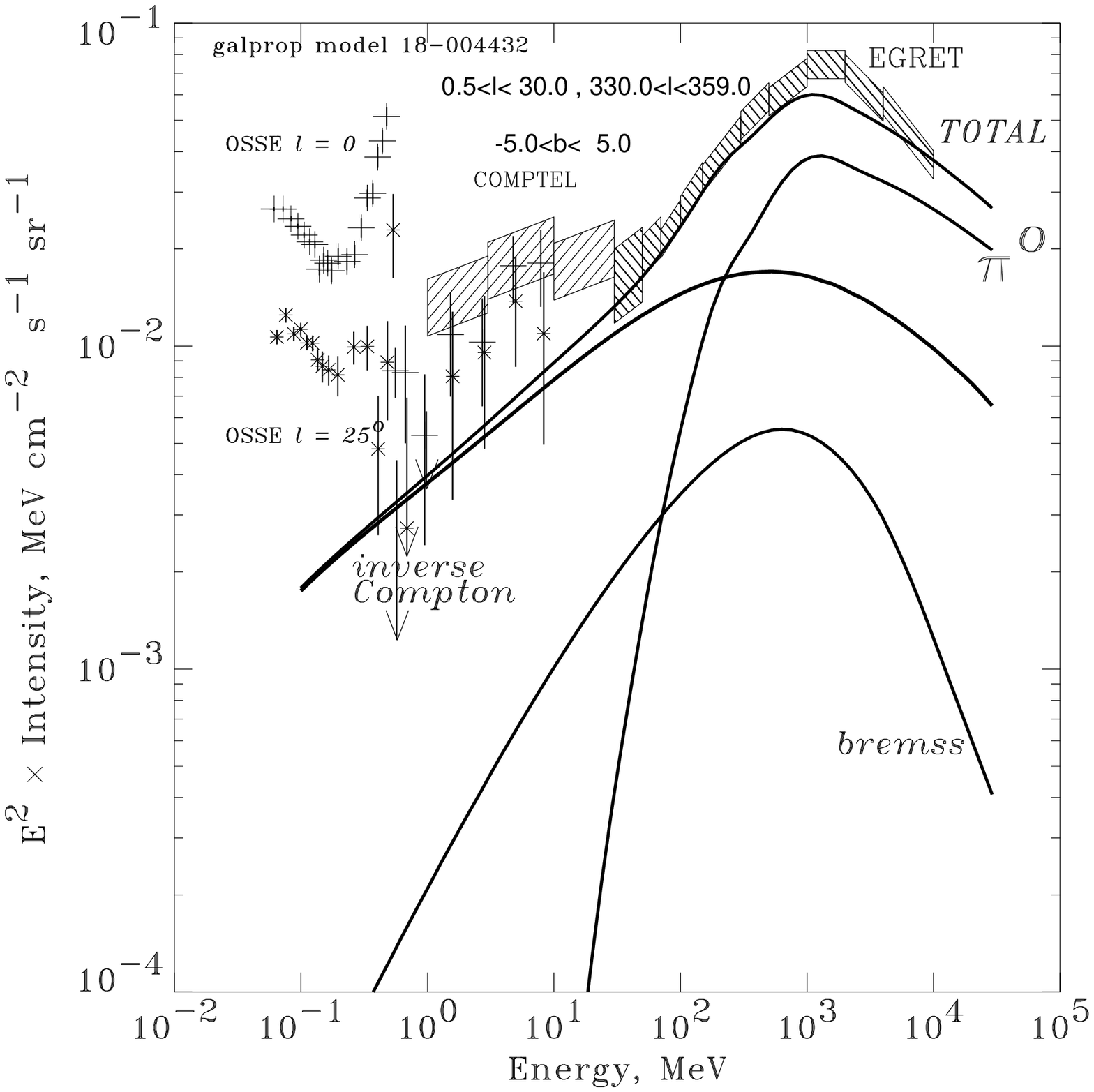,height=3.0in,width=2.8in}
   }
\vspace{10pt}
\caption{Gamma-ray spectrum of inner Galaxy for (left) `conventional' CR
spectra; (right) hard nucleon spectrum.  Data:  OSSE \protect\cite{Kinzer},
COMPTEL \protect\cite{Strong98}, EGRET \protect\cite{StrongMattox96}.
}
\label{gamma_conventional}
\end{figure}

\begin{figure}[] 
   \centerline{\epsfig{file=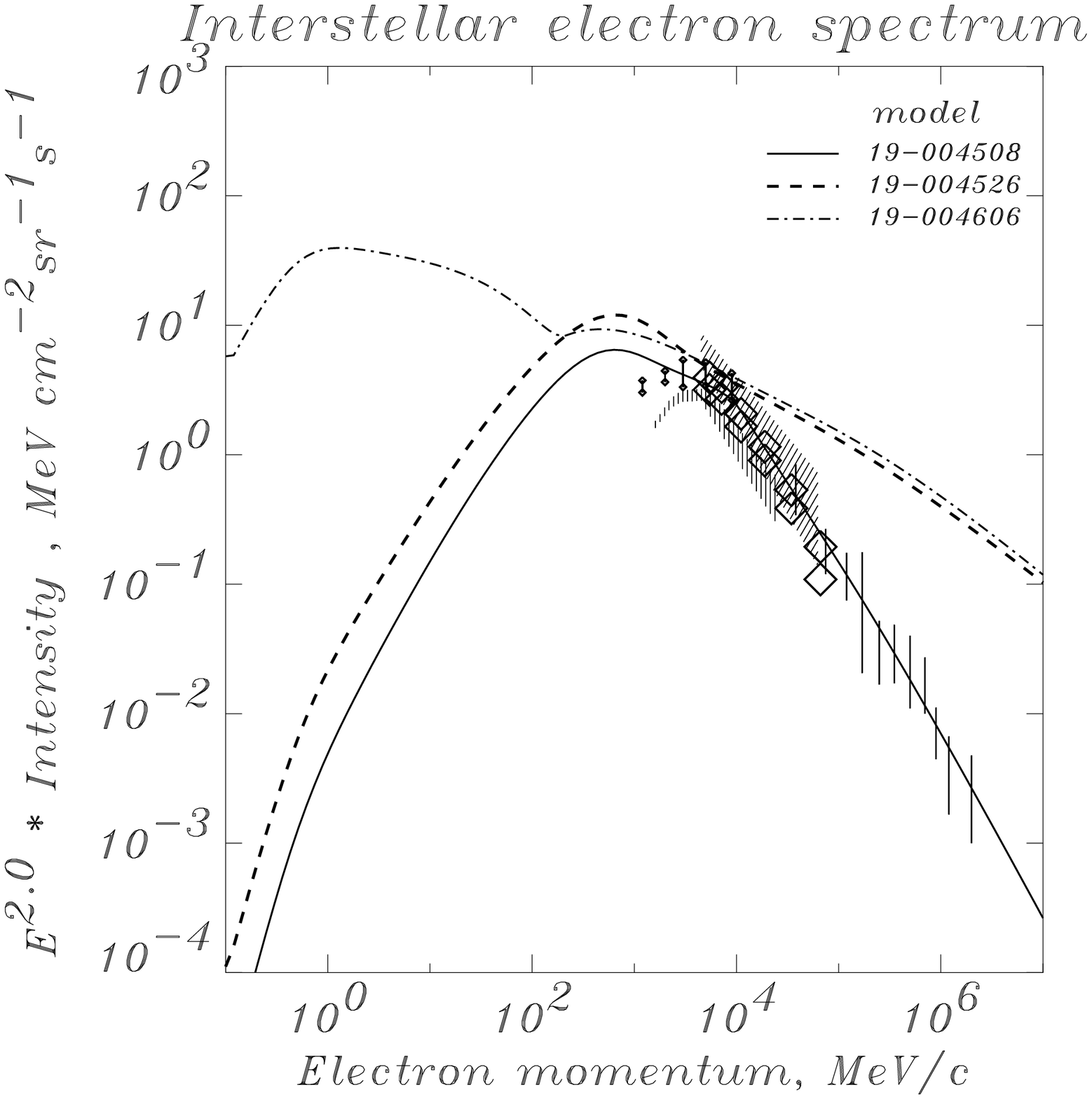,height=3.0in}}
\vspace{10pt}
\caption{Electron spectra observed locally and for various models.  Solid line:
`conventional' model, dashed line:  injection spectrum 1.8, dash-dot line:
spectrum reproducing low-energy $\gamma$-rays.  For data see
\protect\cite{smr98}.
}
\label{electrons}
\end{figure}

\begin{figure}[] 
   \centerline{\epsfig{file=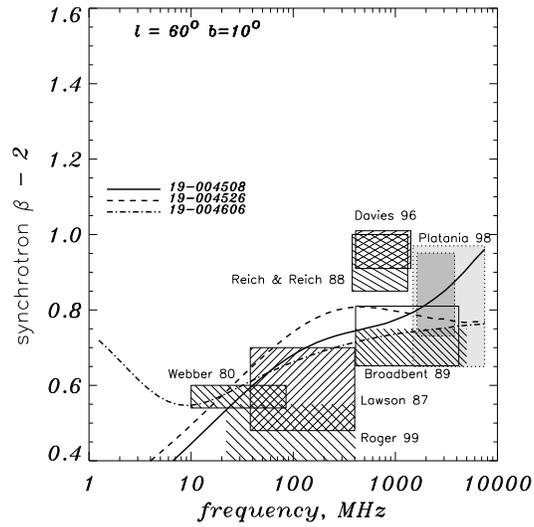,height=3.0in}}
\vspace{10pt}
\caption{Synchrotron index for various electron spectra as in Fig 
\protect\ref{electrons}. For data see \protect\cite{smr98,sm2000}.}
\label{sync_index}
\end{figure}

Figure \ref{gamma_conventional} shows the diffuse spectrum of the inner Galaxy
for what we call a `normal' or `conventional' CR spectrum which is consistent
with direct measurements of high energy electrons and synchrotron spectral
indices (Figs \ref{electrons}, \ref{sync_index}; see \cite{smr98,sm2000}).
Clearly this model does not fit the $\gamma$-ray data at all well.

Consider first the well known problem of the high energy ($>$~1~GeV) EGRET
excess \cite{Hunter97}.  One obvious solution is to invoke $\pi^o$-decay from a
harder nucleon spectrum than observed in the heliosphere, which might for
example be the case if the local nucleon spectrum were dominated by a local
source which is not typical of the large-scale average.  Then the local
measurements would give essentially no information on the Galactic-scale
spectrum.  One can indeed fit the EGRET excess if the Galactic proton (and
Helium) spectrum is harder than measured by about 0.3 in the index (Fig
\ref{gamma_conventional}).

\begin{figure}[b!] 
   \centerline{
            \epsfig{file=fig4a.ps,height=2.5in,width=2.5in}
            \epsfig{file=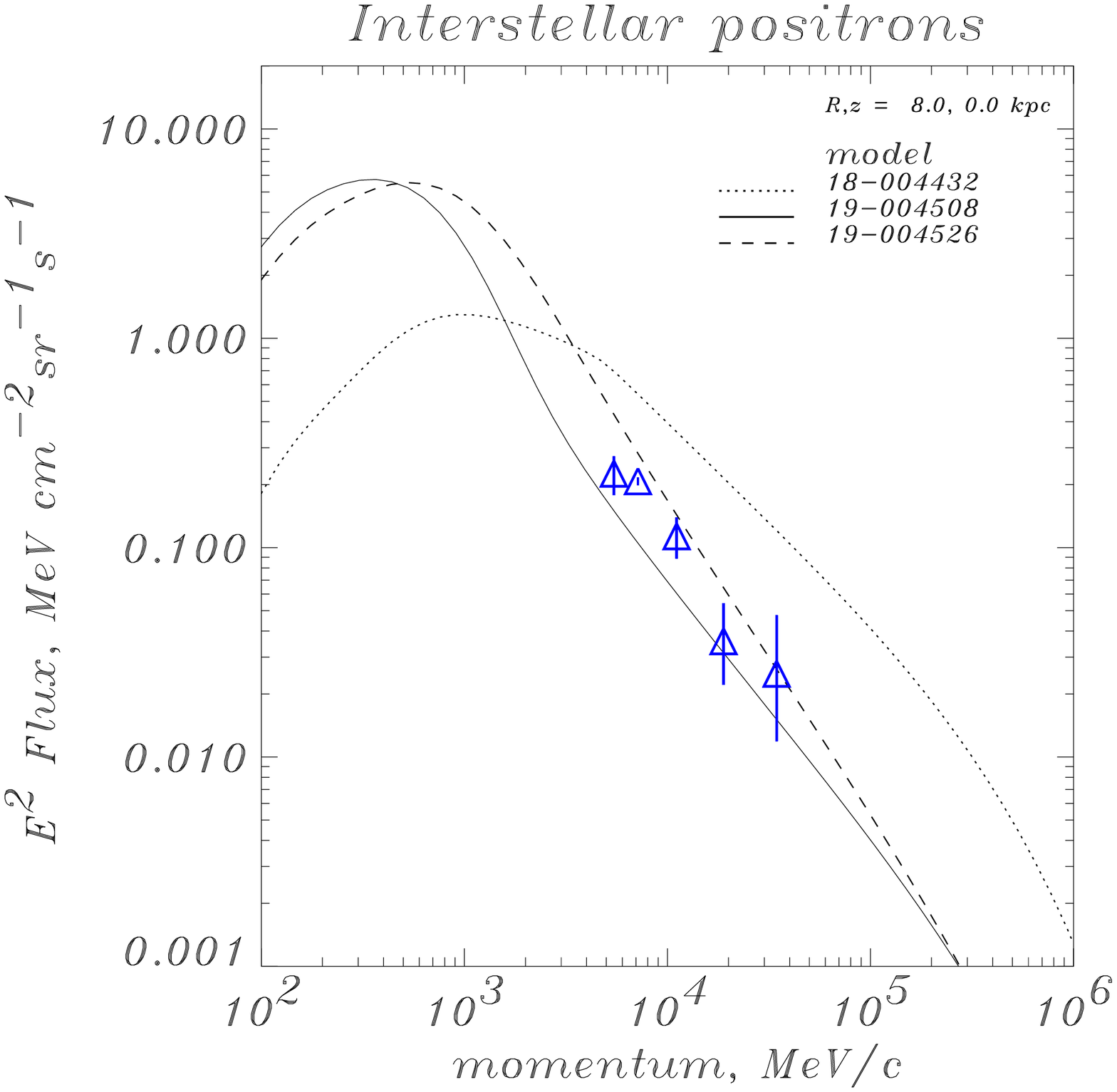,height=2.5in,width=2.5in}
   }
\vspace{10pt}
\caption{Secondary antiproton (left) and positron spectra (right) for a hard 
nucleon spectrum. 
Data: antiprotons \protect\cite{Basini99}, positrons \protect\cite{Barwick}.}
\label{antiprotons_positrons}
\end{figure}

But there are two critical tests of this hypothesis provided by secondary
antiprotons and positrons.  It was shown in \cite{msr98} that such a hard
nucleon spectrum produces too many antiprotons.  The new MASS91 measurements
\cite{Basini99}, which give the absolute antiproton spectrum from 3.7 to 24 GeV,
have clinched this test, as shown in Fig \ref{antiprotons_positrons}.  Quite
independently, secondary positrons give a similar test, which the hard nucleon
hypothesis equally fails (Fig \ref{antiprotons_positrons}).  Again new data,
this time from the HEAT experiment \cite{Barwick}, give a good basis for this
test.  We conclude that there are significant problems if one wants to explain
the GeV excess with $\pi^o$-decay.  This illustrates the importance of
considering all the observable consequences of any model.  Of course it is
anyway difficult to imagine such spectral variations of nucleons given the large
diffusion region and isotropy of CR nucleons.

\begin{figure}[] 
   \centerline{
      \epsfig{file=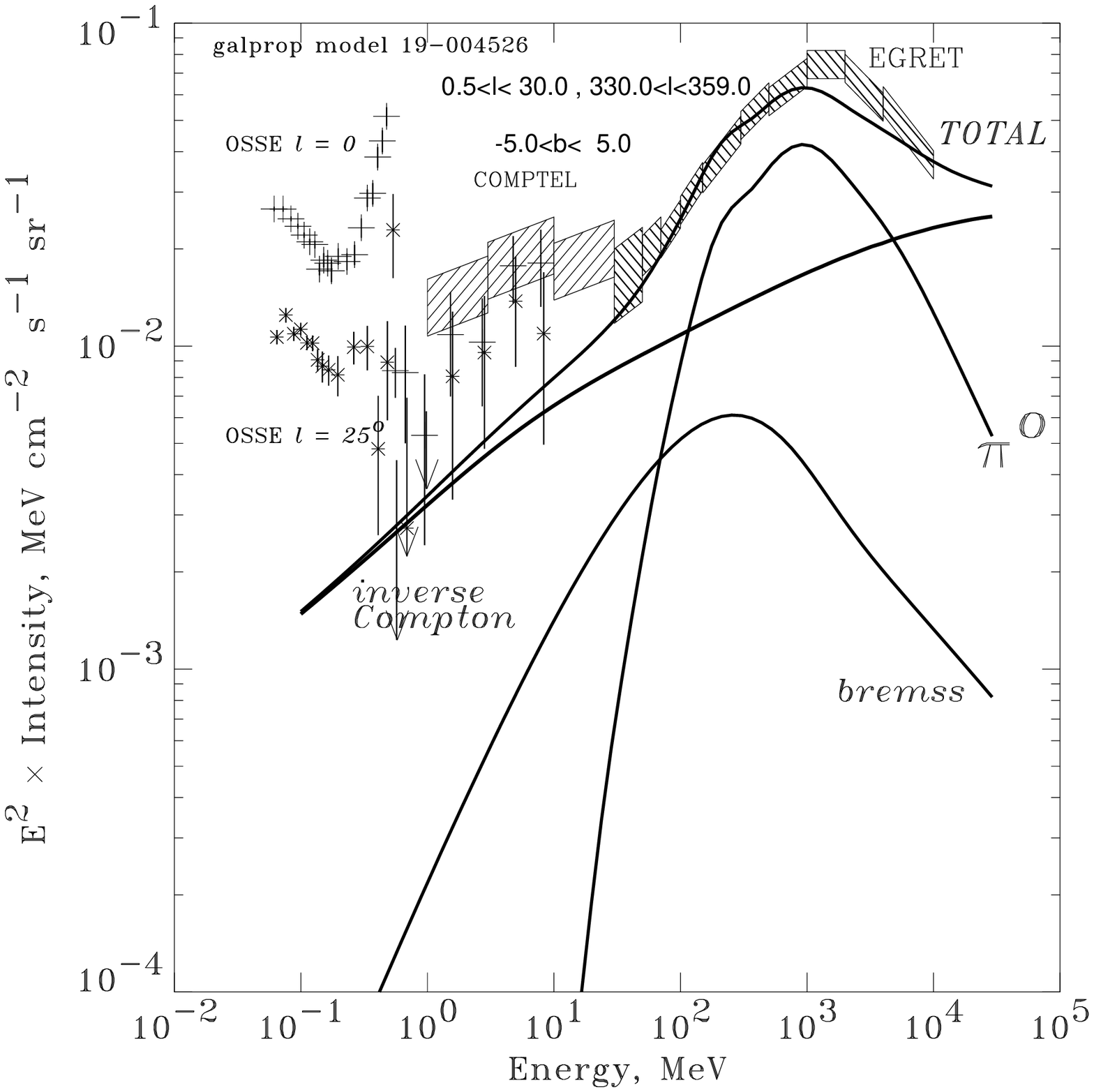,height=2.2in,width=2.8in }
      \epsfig{file=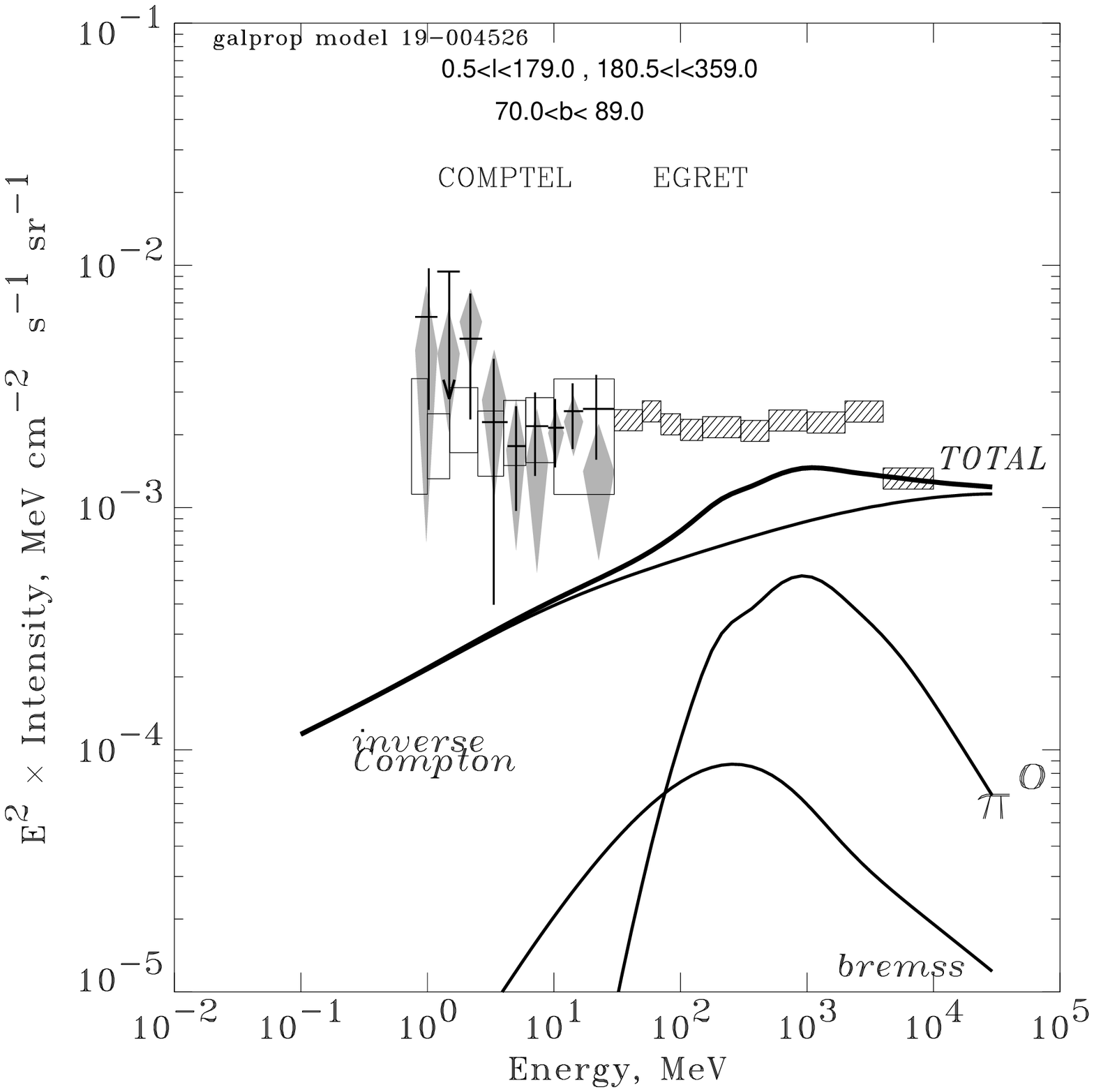,height=2.2in,width=2.8in }
   }
\vspace{10pt}
\caption{Gamma-ray spectrum for a hard electron injection spectrum.  Left:
inner Galaxy ; Right:  high latitudes.  Data as Fig
\protect\ref{gamma_conventional}.
}
\label{gamma_ray_spectra}
\end{figure}

An alternative idea, first investigated in detail in \cite{PohlEsposito}, is
inverse Compton (IC) from a hard electron spectrum.  The point is that the
electron spectrum we measure locally may not be representative of the
large-scale Galactic spectrum due to the large spatial fluctuations which arise
because of the large energy losses at high energies.  What is measured directly
may therefore depend only on the chance locations of the nearest electron
sources, and the average interstellar spectrum could be very different, in
particular it could be much harder.  An injection spectral index around 1.8 is
required (Fig \ref{electrons}) and the corresponding $\gamma$-ray spectrum is
shown in Fig \ref{gamma_ray_spectra}.  Note that modern theories of SNR shock
acceleration can give hard electron injection spectra \cite{Baring} so such a
behaviour is not entirely unexpected.

To predict reliably the IC emission, we also need an updated model for the
interstellar radiation field; we have recomputed it \cite{smr98} using new
information from IRAS, COBE, and stellar population models.  There is still much
scope for further improvement in the ISRF calculations however.

\begin{figure}[] 
   \centerline{
      \epsfig{file=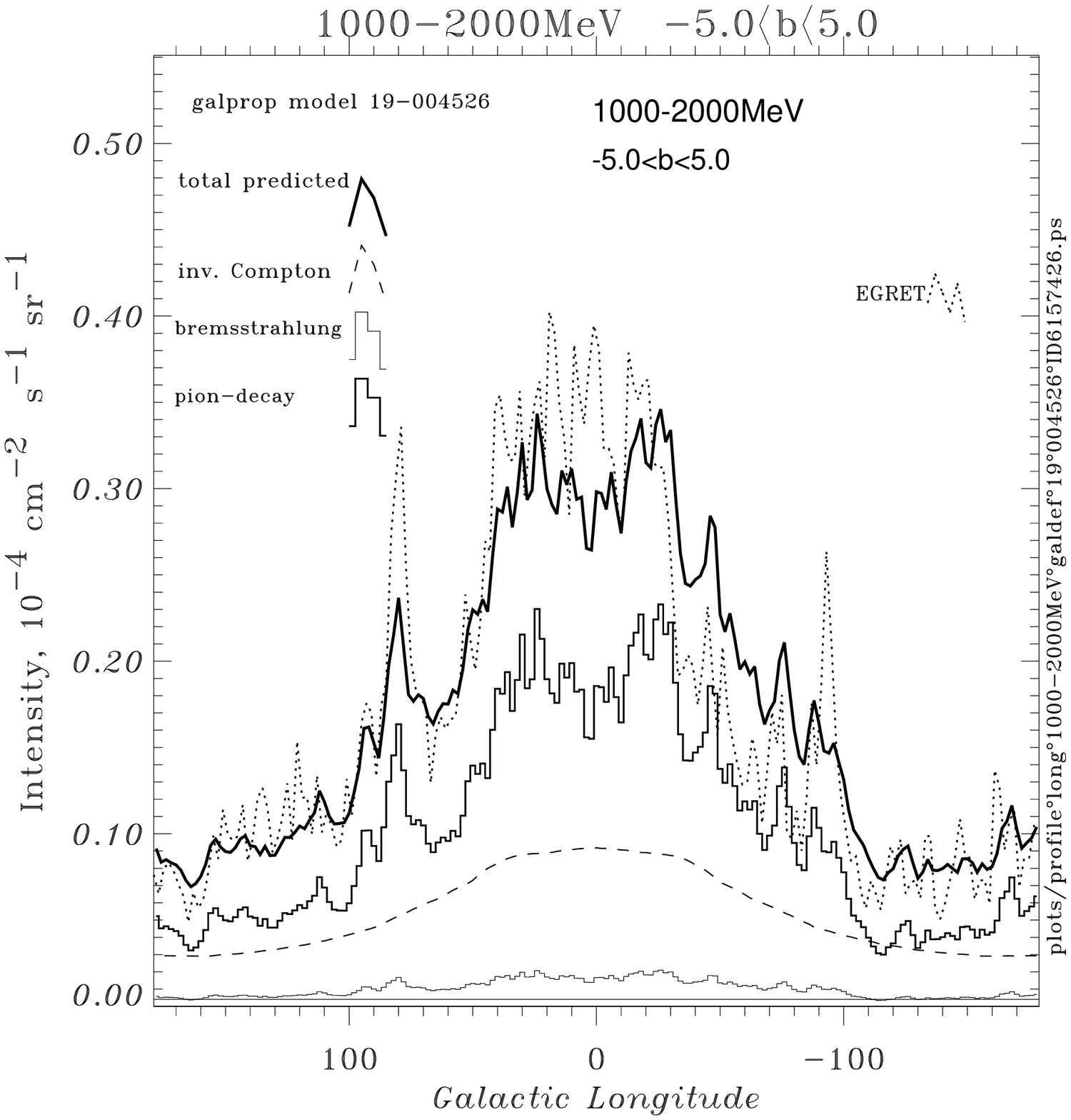,height=2.0in,width=2.5in }
      \epsfig{file=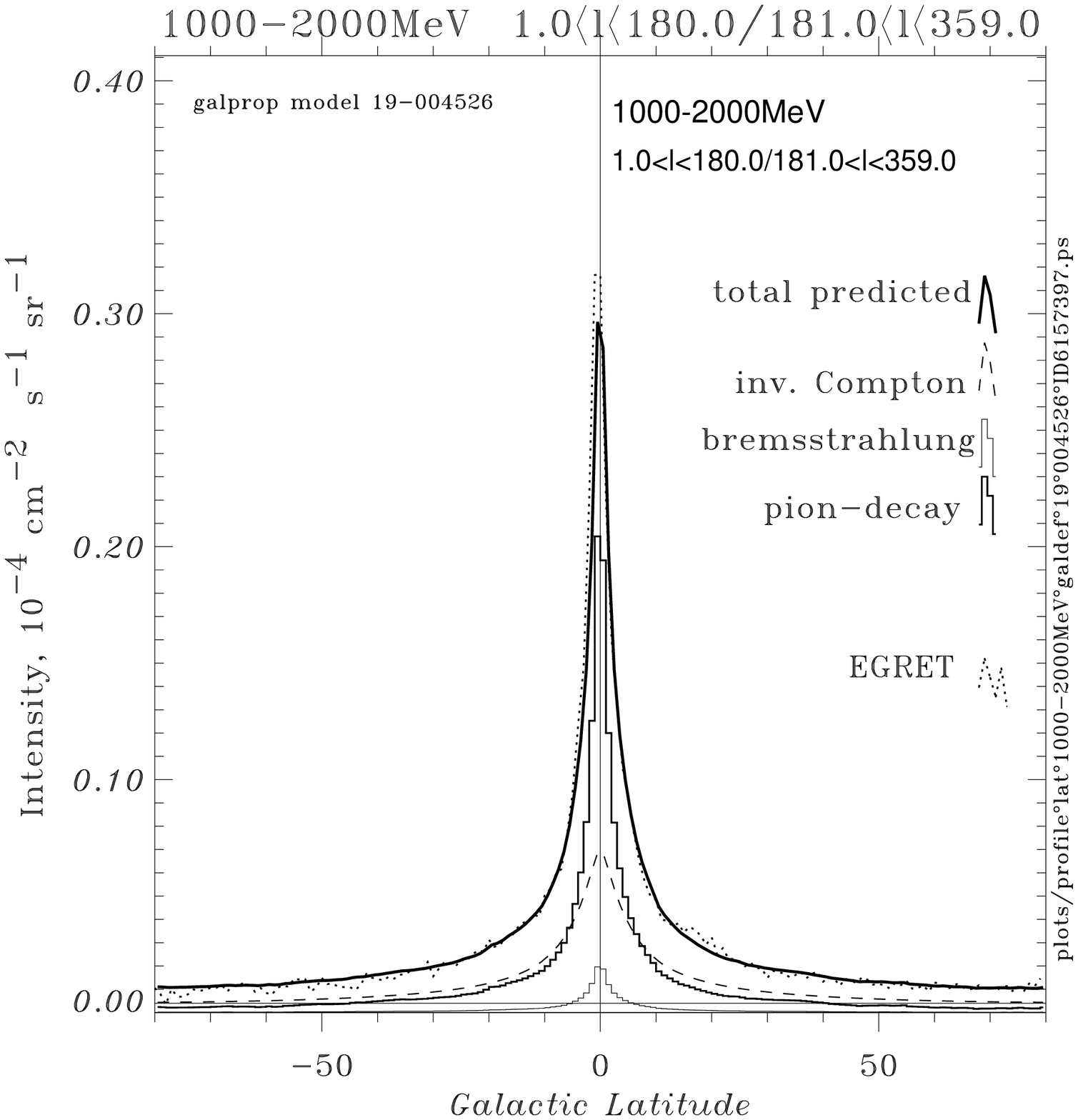,height=2.0in,width=2.5in }
   }
\vspace{10pt}
\caption{Gamma-ray profiles in the energy range 1--2 GeV for a model with a hard
electron spectrum \protect\cite{smr98}.  Dotted line:  EGRET data, dashed line:
inverse Compton, upper histogram:  $\pi^o$-decay, lower histogram:
bremsstrahlung, upper solid line:  sum of components.  Left:  longitude profile,
right:  latitude profile.
}
\label{gamma_ray_profiles}
\end{figure}

\begin{figure} 
   \centerline{
      \epsfig{file=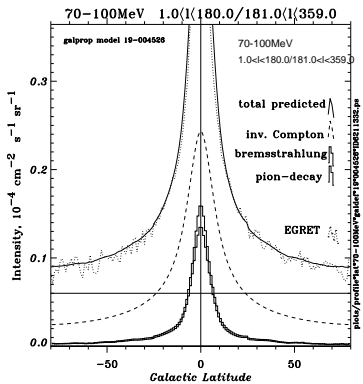,height=2.0in }
   }
\vspace{10pt}
\caption{Gamma-ray profile at high latitudes, for the energy range 70--100 MeV
\protect\cite{smr98}.  Components as Fig \protect\ref{gamma_ray_profiles};
horizontal line:  isotropic background.
}
\label{gamma_ray_high_latitude_profile}
\end{figure}

Note that for these hard electron spectra IC dominates above 1 GeV, and is
everywhere a very significant contributor, while bremsstrahlung is relegated to
third position in contrast to the more conventional picture (presented e.g.\  in
\cite{Strong97}).  Even if we can fit the inner Galaxy spectrum, the critical
test is the spatial distribution:  from Fig \ref{gamma_ray_profiles} one can see
that it can indeed reproduce the longitude and latitude profiles.  In fact it
can reproduce latitude profile up to the Galactic pole (Fig
\ref{gamma_ray_high_latitude_profile}) which is not the case for models with
less IC.  This can be seen as one proof of the importance of IC.  But there is
at least one problem associated with the hard electron spectrum hypothesis.  A
recent reanalysis of the full EGRET data for the Orion molecular clouds
\cite{Digel} determined the $\gamma$-ray emissivity of the gas, and this also
shows the GeV excess, which would not expected since it should not involve IC.
This could be a critical test.  Perhaps the increased radiation field in the
Orion star-forming region could boost the IC, and this ought to be investigated
in detail.

An earlier analysis correlating EGRET high-latitude $\gamma$-rays with 408 MHz
survey data \cite{Chen} found evidence for IC with an $E^{-1.88}$ spectrum.
This is very much in accord with the present models.  More recently a study
\cite{Dixon} which used a wavelet analysis to look for deviations from the
Hunter et al.  \cite{Hunter97} model provided evidence for a $\gamma$-ray halo
with a form similar to that expected from IC.

 An effect which may be important at high latitudes is the enhancement due to
the anisotropy of the ISRF and the fact that an observer in the plane sees
preferentially downward-travelling electrons due to the kinematics of IC
\cite{ms2000}.  This can enhance the flux by as much as 40\% for a large halo.
Even in the plane it can have a significant effect.  Note that the halo sizes
considered here imply an increased contribution from Galactic emission at high
latitudes, which will affect determinations of the isotropic extragalactic
emission.  More precise evaluation of these implications is in progress.

We mention finally low energies, for which a detailed account is given in
\cite{smr98,sm2000}.  Conventionally one invoked a soft electron injection,
$E^{-2.1}$ or steeper, and this could then explain the 1--30 MeV emission as the
sum of bremsstrahlung and IC.  However it seems impossible to find an electron
spectrum which reproduces the $\gamma$-rays without violating the synchrotron
constraints, unless there is a very sharp upturn below 200 MeV; but even there
it fails at to give the intensities measured by OSSE below 1~MeV.  Therefore a
source contribution appears to be the most likely explanation.

\end{document}